# Interfacial solvation explains attraction between like-charged objects in aqueous solution

Alžbeta Kubincová[1], Philippe H. Hünenberger[1] & Madhavi Krishnan[2]*

[1]Laboratory of Physical Chemistry, Department of Chemistry and Applied Biosciences, ETH Zurich, Vladimir-Prelog-Weg 2, CH-8093 Zürich, Switzerland

[2]Physical and Theoretical Chemistry Laboratory, Department of Chemistry, University of Oxford, South Parks Road, Oxford OX1 3QZ, United Kingdom

**Over the past few decades the experimental literature has consistently reported observations of attraction between like-charged colloidal particles and macromolecules in solution. Examples include nucleic acids and colloidal particles in bulk solution and under confinement, and biological liquid-liquid phase separation. This observation is at odds with the intuitive expectation of an interparticle repulsion that decays monotonically with distance. Although attraction between like-charged particles can be theoretically rationalised in the strong-coupling regime, e.g., in the presence of multivalent counterions, recurring accounts of long-range attraction in aqueous solution containing monovalent ions at low ionic strength have posed an open conundrum. Here we show that the behaviour of molecular water at an interface – traditionally disregarded in the continuum electrostatics picture – provides a mechanism to explain attraction between like-charged objects in a broad spectrum of experiments. This basic principle will have important ramifications in the ongoing quest to better understand intermolecular interactions in solution.**

Water is an asymmetric molecule with a strong permanent dipole whose response to an electric field gives the bulk fluid its characteristically high relative dielectric permittivity of about 80. In the absence of an external field random thermal reorientation causes the molecular dipole moment to average out to zero resulting in no net polarisation. However, at an interface in solution, e.g., a cavity, neutral molecule or macroscopic surface, the hydrogen-bonding symmetry is broken, and molecular water is no longer isotropically oriented. In fact this broken symmetry in interfacial orientation is not limited to water and represents a general phenomenon related to the charge-shape asymmetry of the molecule[1]. For water, the bent-core molecular structure and the resulting orientational preference at an interface are commonly invoked to explain thermodynamic phenomena such as the preferential hydration of anions compared to cations[1], ion specific effects on surface tension[2], reduced hydration repulsion between surfaces[3], and crystallisation of charged nanoparticles[4]. Focusing on the interaction between a pair of objects in solution, it seems plausible that any distance dependent alteration in the orientation behaviour of interfacial solvent molecules could be accompanied by a substantial free energy contribution to the potential of mean force. Such a contribution is not accounted for within the framework of continuum-electrostatics theory which regards the solvent a smooth, featureless medium, and could carry profound implications for the interpretation of experimental observations.

**Continuum electrostatics model for the interaction between like-charged particles in solution**

We consider the interaction of two identical like-charged spheres in an aqueous electrolyte containing exclusively monovalent salt at low ionic strength. In low concentrations of monovalent salt (<1 mM) Poisson-Boltzmann (PB) theory provides an accurate description of electrostatic interactions and generally predicts a monotonically increasing repulsion with decreasing interparticle separation[5,6]. Over the past few decades however several independent studies have reported long-ranged attractive interactions between like-charged dielectric particles in low ionic strength solution that depart qualitatively from the PB picture[7-19]. Note that under the relevant experimental conditions, corrections such as those arising from ion correlations, finite ion size and charge density fluctuations are not sufficient to render the screened repulsion attractive at long range[20-22]. The problem has thus far evaded satisfactory explanation and continues to attract great theoretical interest[23].

To understand how the properties of interfacial water molecules and an associated free-energy change may induce a long-range attraction between like-charged particles it is necessary to consider the mechanism of charge regulation. Electrical charge on an object in solution generally arises due to chemical groups via an associative or a dissociative mechanism. For example, acidic groups dissociate to produce an anion and a free proton via an equilibrium reaction $HA \rightleftharpoons H^+ + A^-$ which is governed by an equilibrium constant of dissociation, $K$ and the p$H$ in bulk solution. p$K$, the negative decadic logarithm of $K$, is directly related to the free energy change of the ionisation process and includes a gas-phase component along with the solvation free energies of the reactants and products[24]. For particles carrying ionisable surface groups at a number density $\Gamma$, the net electrical charge density is given by

$$\sigma = z\alpha\Gamma e \quad (1)$$

where $z = \pm 1$ denotes the sign of charge of the ionised group (e.g., $z = -1$ for an acidic group) and $e$ is the elementary charge. $\alpha$, the ionisation probability is in turn given by[25,26]

$$\alpha = \frac{1}{1+10^{z(\mathrm{p}H-\mathrm{p}K)}\exp(z\psi_s)} \quad (2)$$

Here $\psi_s$ represents the value of the dimensionless electrical potential, $\psi = \frac{e\phi}{k_BT}$ at the surface of the particle, and the energy scale $k_BT$ is the product of Boltzmann's constant and the absolute temperature, $T$. Equation (1) serves as the boundary condition for the charge density at the particle surface where the electrical potential that develops in the electrolyte bulk is determined by the non-linear PB equation: $\nabla^2\psi = \kappa^2\sinh\psi$. Here $\kappa = \sqrt{\frac{2e^2c_0}{\varepsilon\varepsilon_0 k_BT}}$ is the inverse of the Debye length - a measure of the distance over which the electrical potential decays from its surface value $\psi_s$, due to screening by the cloud of oppositely charged counterions in solution, $\varepsilon$ is the relative permittivity of the electrolyte medium ($\varepsilon = 78.5$ for water at room temperature), and $\varepsilon_0$ is the permittivity of free space. Thus the ionisation state of the surface groups is indirectly also coupled to the salt concentration, $c_0$ in solution. This study deals with interparticle interactions measured in deionised water of resistivity 18 MΩcm containing dissolved $CO_2$ from ambient air which results in Debye screening lengths, $\kappa^{-1} \sim 100$nm.

It is evident from equation (2) that p$K$ and bulk p$H$ remaining constant, changes in the magnitude of the electrical potential, $\psi_s$ alter the value of the surface charge. This is exactly what happens when two charged objects approach one another. At a distance of a few Debye lengths, each surface is subject to the decaying tail of the electrical potential due to the approaching object. For like-charged entities the magnitude of the surface potential on both particles increases while that of the surface charge decreases, in accord with equation (2). This well-established phenomenon is referred to as "charge regulation", and the electrostatic free energy of the regulated interaction is generally smaller in magnitude than that for surfaces interacting at constant charge[27]. But for all practical purposes the interaction remains monotonically repulsive as long as the particles carry some non-vanishing electrical charge[28].

**Excess hydration free energy of an interface based on molecular simulations**

Molecular Dynamics (MD) studies on neutral cavities in water have shown that interfacial molecules do not orient isotropically as in the bulk but rather exhibit a slight preferential orientation of the negative O atom towards the cavity surface[1]. As illustrated in Fig. 1, the effects of introducing a positive or a negative charge in the cavity are qualitatively different. As a neutral cavity acquires an increasing positive charge the local electric field reinforces the alignment of the interfacial molecules, i.e., strengthens the preferential orientation of the oxygen atoms towards the surface. In contrast, when the cavity acquires an increasing negative charge, the negative oxygen atoms are repelled. The preferential orientation trend observed at the neutral cavity weakens initially, and then inverts: the water molecules flip around, going through a vanishing preferential orientation at some critical value of negative charge. Beyond this point, the positive hydrogen atoms point preferentially towards the cavity surface, and this trend in orientation is reinforced as the magnitude of negative charge further increases. Similar observations have been made in simulation studies involving planar interfaces [2,3,29,30].

In order to quantify the free energy of interfacial water at a charge-regulating surface we performed MD simulations of pure water at a planar interface. We set up a parallel plate capacitor composed of two walls of area 100 nm$^2$ carrying equal and opposite charge, separated by a gap of about 4 nm (Fig 2a). The plates consisted of fixed hexagonally packed non-polar and non-hydrogen-bonding atoms of the size of the oxygen atom. The gap was filled with 12448 simple point charge (SPC) water molecules. Atoms at the plate surface are assigned a charge of either 0 or $\pm 1\ e$ so as to

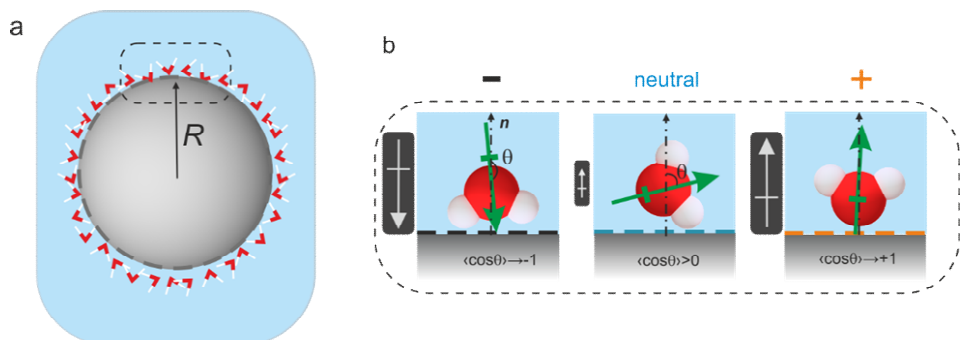

**Fig. 1**. **Molecular water at a charged interface.** a, Schematic depiction of water molecules at the interface between a particle of radius, $R$ and aqueous solution. b, Configurations illustrative of the average orientation of interfacial water molecules at surfaces that carrying a strongly negative charge (left), no charge (centre), or a strongly positive charge (right). $\theta$ is the angle included between the molecular dipole moment (green) and the outward pointing surface normal, $\boldsymbol{n}$ (dashed line). Molecular orientation is inferred from Molecular Dynamics (MD) simulations (Fig. 2b), with $\langle \cos\theta \rangle$ large and negative ($\rightarrow$ -1) for the strongly negative surface, large and positive ($\rightarrow$ +1) for the strongly positive surface, and slightly positive (>0) for the neutral surface. This implies that as the surface charge changes from zero to strongly negative, there is an inversion in the average orientation of interfacial water.

attain final charge densities of $+|\sigma|$ and $-|\sigma|$ at the left and right plates respectively. Polarisation profiles $P(z)$ as a function of the distance $z$ from the positive plate were calculated based on 5 ns simulations for different values of $|\sigma|$ (Fig. 2b). We note that within a region of about 0.5 nm from each surface the polarisation departs substantially

7from the value $\frac{\varepsilon-1}{\varepsilon}\sigma$ expected for an ideal dielectric medium of relative permittivity $\varepsilon$. In order to quantify the excess hydration free energy, i.e. the free energy associated with this extra non-dielectric polarisation component, the continuum value was subtracted from $P(z)$ and the resulting function was integrated from a reference position at the midplane of the capacitor, $z_{\text{mid}}$ up to the surface of each plate (Fig. 2a, Supplementary Section 1). This gives the excess electrical potential $\phi_{\text{cap}}$ at the surface of each plate as a function of charge density, $\sigma$ (Fig. 2c). A charging integral of the form $f(\sigma) = \int_0^\sigma \phi_{\text{cap}}(\sigma)\mathrm{d}\sigma$ then gives the excess hydration free energy per unit area as a function of $\sigma$ due to the excess polarisation (Fig. 2d). The form of $f(\sigma)$ agrees with the qualitative considerations presented in the context of Fig 1. For a positively charged surface, $f(\sigma)$ decreases monotonically with increasing charge density. In contrast for a negative surface, $f(\sigma)$ is a non-monotonic function of $\sigma$. $f(\sigma)$ increases from $\sigma = 0$, goes through a maximum at $\sigma \approx -0.3\ e/\text{nm}^2$ and decreases thereafter.

This reorientation effect for negative surfaces is supported by independent experimental evidence. The orientation of water at charged interfaces has been extensively studied using non-linear optical spectroscopy[31] on various types of surface e.g., silica, alumina, positive and negative lipid bilayers. These studies have shown that water molecules are strongly oriented at charged interfaces due to charge-dipole interactions and are only weakly oriented at neutral surfaces[32-36]. Importantly, sum-frequency generation vibrational spectroscopy on negatively charged silica surfaces in water have reported that not only is molecular orientation at an interface a function of surface charge density but that water molecules indeed flip around with increasing p$H$[34,35,37-39] – which corresponds to increasing negative charge – in line with the behaviour suggested by MD simulations (Figs. 1 and 2).

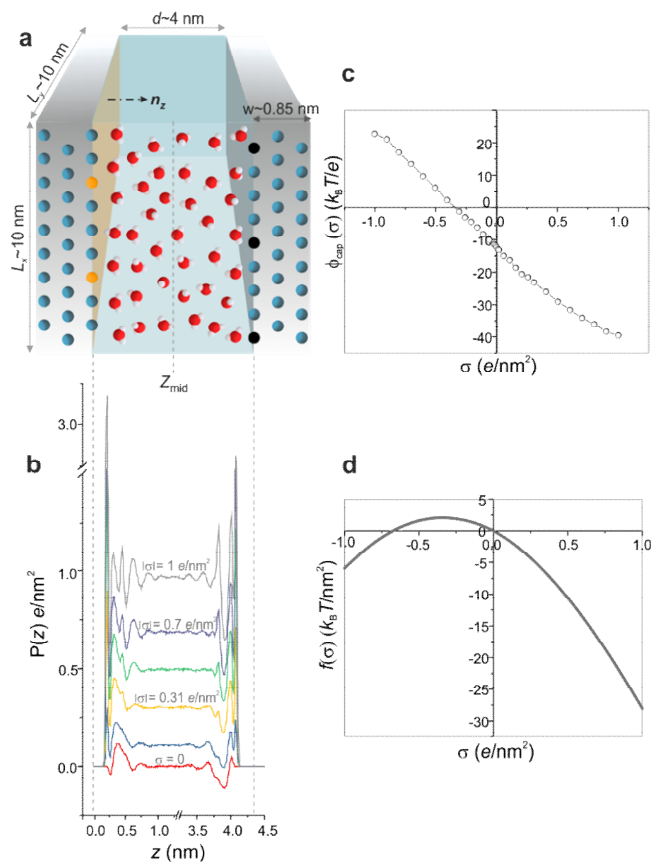

**Fig. 2**. **Excess hydration free energy of a charged interface based on molecular simulations.** a, Schematic representation of the simulated system. The 10 x 10 nm$^2$ parallel-plate capacitor is made up of a positive (left) and negative (right) plate, each composed of three layers of atoms, separated by a gap of about 4 nm filled with water molecules. Interfacial atoms carry a charge of 0 or $\pm 1$ $e$ to attain an overall charge density of $\pm\sigma$ $e$/nm$^2$. The arrow depicts the axis, $\boldsymbol{n}_z$. b, The area-averaged profile, $P(z)$ of the polarisation projected along the $z$-axis is extracted from the simulation. Note that the projection along $\boldsymbol{n}_z$ corresponds to a projection along the outward-directed surface normal $\boldsymbol{n}$ for the left plate (i.e., as in Fig 1b), but an inward-directed normal $-\boldsymbol{n}$ for the right plate (i.e., opposite to Fig 1b). Within about 0.5 nm from the surfaces, the $P(z)$ profiles differs substantially from the bulk value. c, The electrical potential, $\phi_{\text{cap}}(\sigma)$ at the walls as a function of $\sigma$ is derived from $P(z)$ by integration from the reference position, $z_{\text{mid}}$. (d) Based on the $\phi_{\text{cap}}(\sigma)$ curves, the excess hydration free energy $f(\sigma)$ per unit area is derived by integration over $\sigma$ (see Supplementary Section 1).

**A model of interparticle interactions including the excess hydration free energy of the interface from molecular simulations**

We now incorporate the results from molecular simulations into a calculation of the potential of mean force for the interaction of two particles in solution. We represent the total free energy as

$$\Delta F_{\text{tot}}(x) = \Delta F_{\text{el}}(x) + \Delta F_{\text{int}}(x) \quad (4)$$

where $x$ is the inter-surface separation between the particles and each term denotes a free-energy difference with reference to the zero-point set at infinite separation (Fig. 3a). We solve the PB equation for the electrical potential, $\psi$ in the electrolyte region between two particles of radius $R$ at a variable separation $x$, using equation (1) as the boundary condition on all surfaces. We then evaluate the electrostatic interaction energy using a combination of volume and surface integrals as described previously[26,40-42]. Thus we have

$$F_{\text{el}} = -\int_V \left\{ \frac{\varepsilon \varepsilon_0}{2} \boldsymbol{E} \cdot \boldsymbol{E} + 2c_0 k_B T (\cosh \psi - 1) \right\} dV + \Gamma k_B T \int_S \ln \frac{1-\alpha(x)}{1-\alpha(\infty)} dA \quad (5)$$

where $\boldsymbol{E}$ denotes the electric field. The interfacial term, $F_{\text{int}}(x)$ in equation (4) represents the contribution from the orientational behaviour of the interfacial water molecules. This term is calculated based on the MD simulation results for the excess hydration free energy $f(\sigma)$ per unit area (Fig. 2d). Owing to charge regulation, the charge density $\sigma$ at any point of the particle surface is a function of the inter-surface separation $x$ (Fig. 3a). Thus, for a given value of $x$, the term $F_{\text{int}}$ is calculated via the surface integral

$$F_{\text{int}}(x) = \int_S f(\sigma(x)) \, dA \quad (6)$$

Note that the assumption of free-energy additivity implicit in equation (4) has a long history. Dating back at least to the Derjaguin-Landau-Verwey-Overbeek (DLVO) theory, such assumptions are widely used in colloid science in order to partition interaction free energies [43,44]. In particular, concerning the summation of hydration and electrostatic forces, the assumption has been explicitly tested in atomistic simulations and found to hold within accuracy limits under the relevant conditions[3].

**Attraction in the weakly-charged negative regime caused by the interfacial free-energy component**

Figure 3 illustrates the proposed mechanism by which an attraction may manifest in the interaction of like-charged particles in solution. As two like-charged objects approach, regulation decreases the magnitude of their electrical charge (Fig. 3a, c), but the counterions in the gap, which are required to preserve electroneutrality, resist compression. Therefore as long as the particles retain a net electrical charge, the overall electrostatic component of the interaction, $\Delta F_{el}$ – including both the field energy and configurational entropy of the ions – generally remains repulsive over the entire distance range[27,40].

But according to the MD results, and independent spectroscopic confirmation, a reduction in surface charge density influences the average orientation of interfacial water and therefore alters the interfacial hydration free energy (Fig. 3b). MD suggests that in the weakly charged regime, $|\sigma| < 0.3$ $e$/nm$^2$, approach of two negative like-charged particles is accompanied by a reduction in solvation free energy of interfacial water, $F_{int}$ owing to a down regulation of surface charge (Fig. 3d). This attractive interfacial contribution counteracts the increase in free energy due to electrostatic repulsion, $F_{el}$. In the regime of finite sized interacting spheres, $\kappa R \leq 50$ say, the attractive interfacial contribution can under certain conditions dominate the electrostatic repulsion and result in an interaction energy minimum in the potential of mean force at fairly long range ($x > \kappa^{-1}$) (Fig. 3e, black curve). Importantly, this model does not envisage an attraction for approaching positive particles, as here the hydration free energy of water increases monotonically with decreasing surface charge. This implies a repulsive rather than an attractive interfacial contribution to the total free energy for approaching positive particles (Fig. 3e, orange curves).

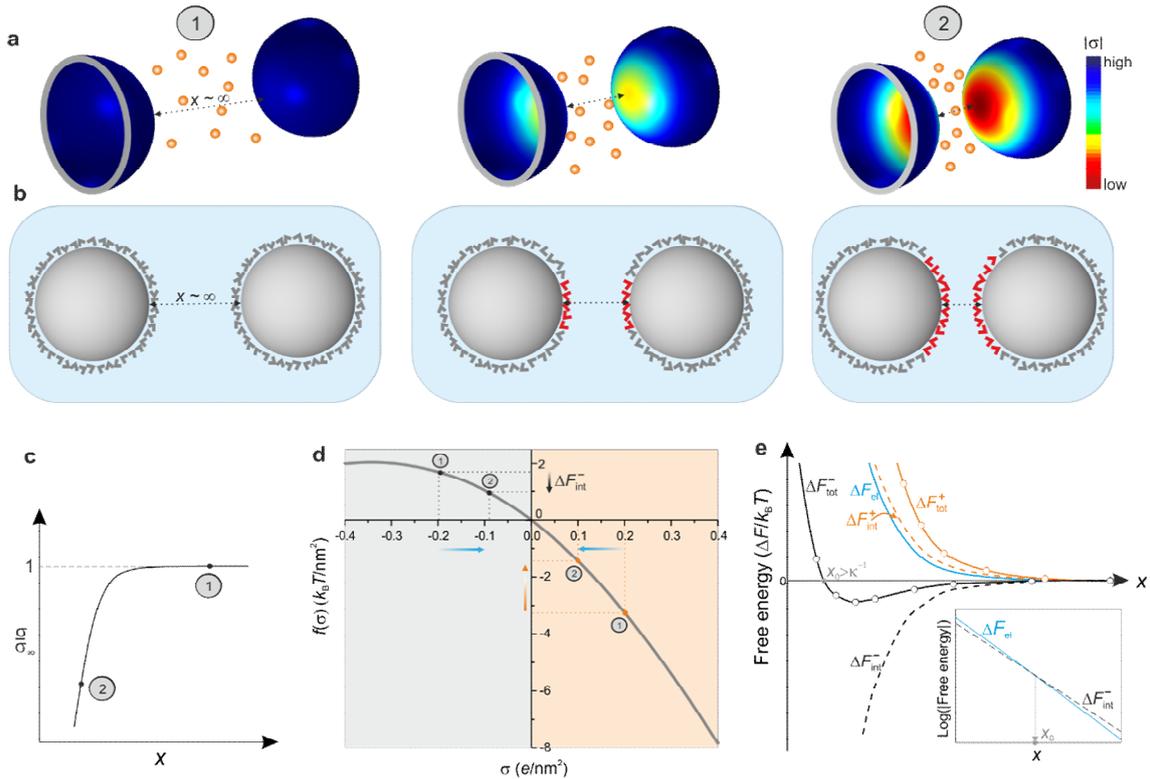

**Fig. 3**. **Mechanism explaining attraction between like-charged objects in solution.**
a, As two like-charged particles approach each other from large separation (left to right) the charge density, $\sigma$ on the facing regions decreases in magnitude due to charge regulation. The counterions in the gap (orange spheres, not to scale) give rise to an entropic repulsion. b, Since the orientation of the interfacial water molecules is a function of surface charge density, (Figs. 1b and 2b), water molecules in the facing regions (coloured red) respond to the local change of $\sigma$. The schematic depiction of water orientation is for weakly charged negative surfaces. c, A schematic representation of charge density, $\sigma$ relative to its value at infinite separation, $\sigma_\infty$ as a function of $x$ for points on the surface lying on the line connecting the particle centres. d, Response of the interfacial solvation energy to a reduction in magnitude of $\sigma$ in the facing regions for both negative and positive particle charge. In the regime $|\sigma| < 0.3$ $e$/nm$^2$, reorientation of the interfacial water molecules results in a reduction in free energy for negative particles but an increase for positive particles. e, The sum of the electrostatic free energy, $\Delta F_{el}$(blue curve) and the water contribution $\Delta F_{int}$(dashed black line) results in a total potential, $\Delta F_{tot}$ (solid black curve) that can go through a minimum at very long range for the interaction of two negative particles, while it remains monotonically repulsive for the interaction of two positive particles (solid orange curve). The inset illustrates how the magnitude of the attractive interfacial contribution $|\Delta F_{int}^-|$ can dominate at long range, while the repulsive $|\Delta F_{el}|$ is dominant at short range. The case presented corresponds to conditions similar to that for curve viii in Fig. 4.

We remark that in practice it is challenging to obtain negatively charged surfaces with $|\sigma| \gg 0.3$ $e$/nm² in electrolytes of $p$H $\leq 7$, particularly at low ionic strength ($c_0 <$ 1 mM) (Supplementary Fig. S1). Even at higher ionic strengths of ca. 100 mM and ionisable group densities $\Gamma > 0.3$ /nm², the regime of $|\sigma| < 0.3$ $e$/nm² is attained for $p$H values up to about one unit higher than the p$K$. For $\Gamma < 0.3$ /nm² however, $|\sigma|$ is always less than 0.3 $e$/nm², and an attractive solvation energy contribution could be relevant regardless of p$H$ and ionic strength (Supplementary Fig. S1 c). This implies a potentially ubiquitous role for interfacial water in interactions between weakly negatively charged objects.

**Comparing the calculated total interaction free energy with experiment**

We now compare our calculations of the total free energy, $F_{\text{tot}}(x)$ with observations from two experimental studies in the literature. The first set of measurements concerns interparticle interaction potentials inferred from radial distribution functions, g($r$) for negatively charged polystyrene latex spheres of radius $R$=0.65 μm in low ionic strength solution measured using optical microscopy (Fig. 4a)[10]. The experimental conditions involved Debye screening lengths $\kappa^{-1} \sim 50 - 250$ nm, and the measurements typically revealed long-ranged attraction interaction potentials with shallow minima located between 0.5 and 2 μm (Fig. 4b-d, square symbols). The electrolyte in the study was deionised water which typically contains ions at a concentration $c_0 \cong 10 \, \mu M$ and has a $pH \approx 5.5$ due to dissolution of $CO_2$ from ambient air (Fig. 4b).

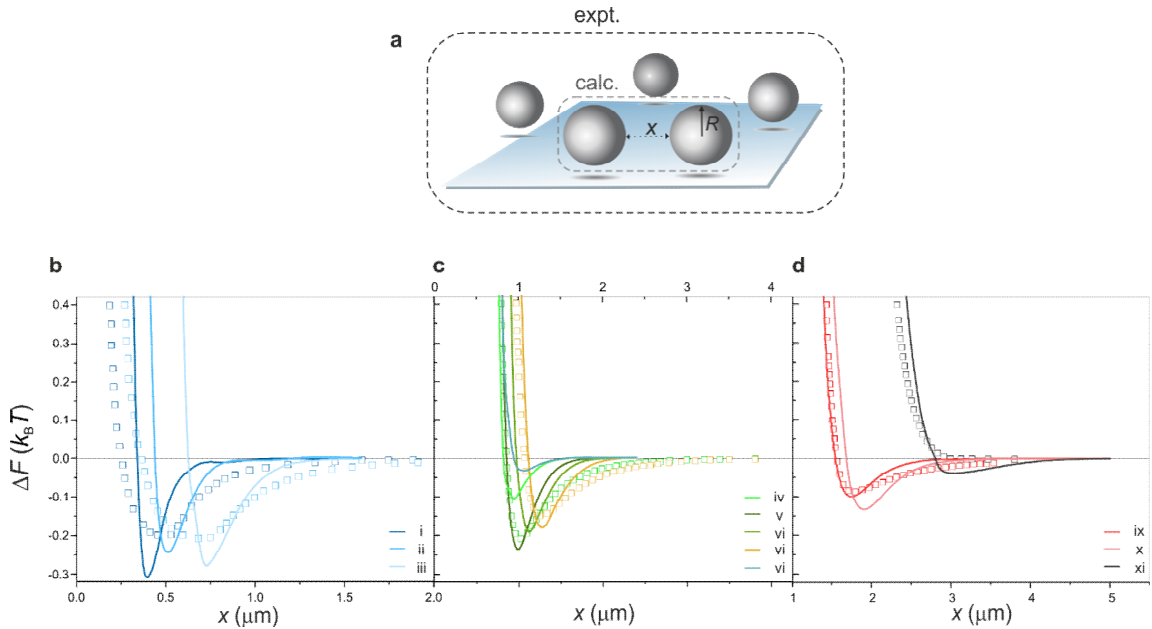

**Fig. 4. Comparison of calculated interaction potentials with experimental data from Ref. 10.** a, Schematic depiction of the experimental situation where the diffusion of an ensemble of polystyrene colloidal particles of radius, $R = 0.65$ μm is monitored above the surface of a glass coverslip using optical microscopy. b-d, Pair potentials of the Lennard-Jones form inferred in the study[10] (square symbols), and our calculated potentials for $\Gamma$=0.1 e/nm$^2$ (solid lines). Parameters for the calculated curves are quoted in the format ($p$, $c_0$ in μM, $\kappa^{-1}$ in nm). b, Measurements in deionised water with estimated $\kappa^{-1}$ values of 62 nm (dark blue symbols) and 92 nm (light blue symbols). Parameters for the calculated curves are i: (-2, 16.6, 75), ii: (-2, 25, 61), and iii: (-2, 8.3, 106). c, Measurements with deionising resin achieving an intermediate ionic strength with estimated $\kappa^{-1}$ values of 140 nm (green symbols) and 170 nm (yellow symbols). Parameters for calculated curves are iv: (-2.1, 6.67, 118), v: (-2.05, 5, 136), vi: (-2.15, 4.17, 149), vii: (-2.15, 3.33, 167), and viii: (-2.2, 6.67, 118). d, Measurements with deionising resin achieving the lowest ionic strength with nominal fit $\kappa^{-1}$ values of 230 nm (red symbols). Parameters for calculated curves, ix: (-2.4, 2.1, 210), x: (-2.45, 1.67, 235), and xi: (-3.8, 0.83, 334).

Contact with deionising resin in some measurements (Figs. 4 c and d) reduces the ion concentration by an order of magnitude down to the level of about $c_0$=1 μM, which corresponds approximately to a resistivity of 18 MΩcm. It is fair to assume that this simultaneously returns the p$H$ of the electrolyte to its neutral value of 7. Since the p$H$ and ion concentration could not be measured directly in the experiments, we work with values of p$H$ and $c_0$ estimated as described above.

The solid lines present calculations of $\Delta F_{\text{tot}}(x)$ based on a nominal surface density of ionisable groups, $\Gamma$ =-0.1 e/nm² in electrolytes of various ionic strengths. Given the uncertainty in the effective p$K$ value of the surface sulfonate groups in equation (2), this quantity was treated as an adjustable parameter. Best agreement between experiment and calculation was obtained for p$K$ between 3 and 3.5. Although the p$K$ of the isolated sulfonic acid is about -0.5, measurements and calculations indeed suggest a value of about 3 for oligomers of styrenesulfonic acid[24]. The proximity of the low-dielectric particle interior as well as the inclusion of small amounts of carboxylic acid groups during the synthesis process may also contribute to a slightly increased effective p$K$. In the figure, calculated pair potentials are shown for various combinations of the parameter $p = z(\text{p}H - \text{p}K)$ from -3.8 to -2 and corresponding $c_0$ values in the range 1-25 μM.

Considering the wide range of qualitatively different behaviours resulting from slightly different calculation parameters (*e.g.,* Fig. 4c), it appears that the experimental observation could range from a nearly vanishing attraction to a minimum in the potential of depth around 0.5 $k_\text{B}T$. The calculations thus suggest that the presence of the well, and its depth, would be highly sensitive to the p$H$ and the ionic strength;this is in line with reports from the experimental literature[14,16,45,46]. In practice, small drifts of conductivity and p$H$ during the measurements, variability in the particle size $R$ and ionisable group density $\Gamma$, along with out-of-plane (vertical) motion of the particles, are all expected to smear out the measured response[10], thereby causing unavoidable discrepancies between experiment and calculation. In particular, the functional form of the measured pair potentials at close approach ($x < 0.5$ μm) could be particularly sensitive to uncertainties in particle size $R$ (~ 0.1 μm). Nevertheless, we obtain remarkable agreement between calculated and experimental curves for plausible values of the system parameters.

**Interfacial hydration explains symmetry-breaking behaviour in like-charge interaction**

The second set of experimental observations concerns reports of Groves *et al.* which extended the like-charge attraction observation to much larger micron-scale silica particles, $R = 3.25$ μm, coated with lipid bilayers composed of a mixture of charged and uncharged lipids with tuneable composition[15,18]. Here the observed long-ranged interparticle attraction is so strong that it results in stable clusters of hexagonally close-packed particles (Fig. 5a, bottom panel), implying attractive minima in the pair potentials whose depth is at least an order of magnitude larger than that the polystyrene latex sphere experiments[10,46]. Again, as previously reported the attractive minima as inferred from the measured radial distribution functions occur at intersurface separations of several hundreds of nanometres (Fig. 5b). However, very intriguingly this study reported attractions only for negatively charged particles and not for particles coated with net positively charged lipid bilayers.

The experiments were performed using mole fractions of charged lipids of 1-5 % for the negative lipids and 7-11 % for the positive case. Assuming an area per lipid head group of 2 nm$^2$, the charge densities probed correspond to ranges in $\sigma$ of

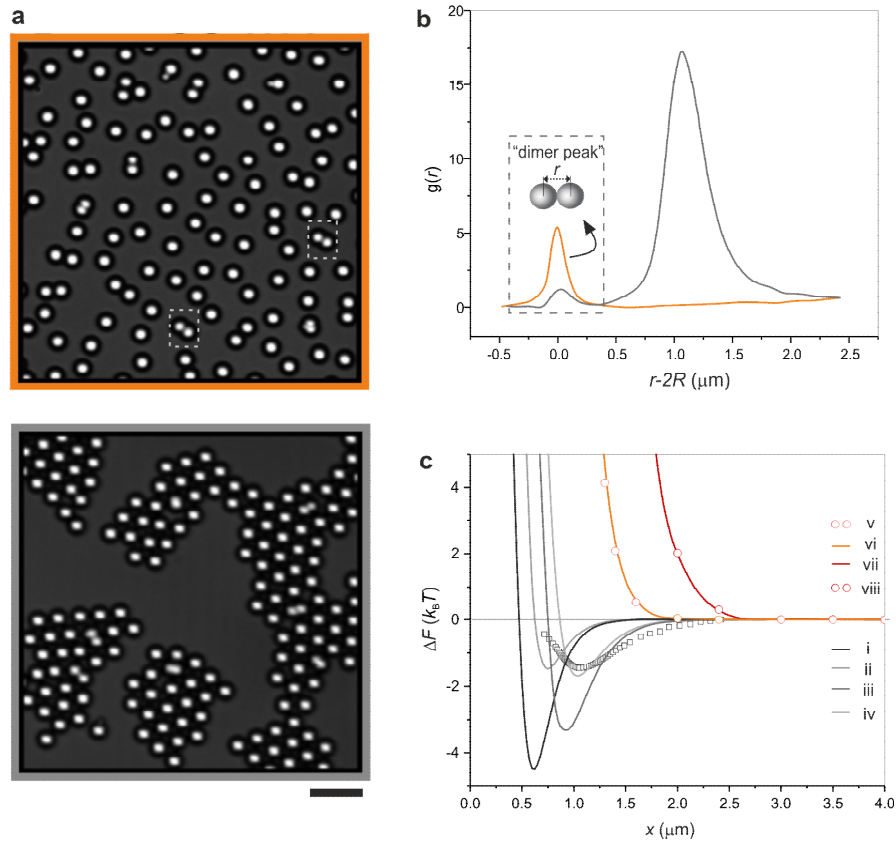

**Figure 5**. **Broken symmetry in the response of the pair potential to a change in sign of particle charge.** a, Optical microscopy snapshots of particles coated with positive lipid bilayers (top) and negative bilayers (bottom) indicating repulsive interactions in the former and long range attractive interactions in the latter case (images reproduced with permission from Ref. 18). Scale bar denotes 20 μm. "Dimers" (demarcated in dashed boxes) are irreversibly aggregated pairs of particles inevitably present in colloidal preparations and irrelevant to the interpretation of long-range interparticle attractions of interest here. b, Radial probability density distributions, $g(r)$ reported for the above cases including only the first significant peak in $g(r)$ (Black curve - negative particles, orange curve – positive particles). c, Calculated interaction potentials for $R = 3.25$ μm particles (solid curves) with parameters presented in the format ($p$, $c_0$ in μM, $\sigma$ in $e$/nm$^2$). Curves for negatively charged particles, i: (-2.3, 4.17, -0.01), ii: (-2.1, 4.17, -0.025), iii: (-2.28, 2.1, -0.01), and iv: (-2, 2.1, -0.025). Curves for positively charge particles, v: (-4.5, 4.17, +0.035), vi: (-4.5, 4.17, +0.055), vii: (-4.5, 2.08, +0.035) and viii: (-4.5, 2.08, +0.055). Salt concentrations, $c_0 \approx 2-4$ μM in the calculations are comparable with the reported experimental value of $c_0 \sim 5$ μM for deionised water equilibrated with air. In the absence of a measured pair potential, $-\ln g(r)$ is reported as a crude estimate (square symbols, based on the black curve in panel b) of the range and depth of the attractive minimum for the experiments on negative particles.

-0.005 to -0.025 $e$/nm$^2$ and +0.035 to +0.055 $e$/nm$^2$. Calculations of $\Delta F_{\text{tot}}(x)$ reveal deep minima of about 1-4 $k_\text{B}T$ at experimentally reported interparticle distances of around 1 μm for the negatively charged system (curves i-iv, Fig. 5c). Given that the p$H$ of water exposed to air is ~5.5-6, the values of $p$ in the calculated curves imply p$K$ values of 3.2-3.9 which are in excellent agreement with the reported p$K$ values of negatively charged lipid head groups used in the work[47]. The significantly larger well depths in these experiments with $R = 3.25$ μm compared with the polystyrene experiments where $R = 0.65$ μm (Fig. 4) would be consistent with an attractive contribution growing approximately in proportion to the particle surface area (ratio of areas = 25). Such scaling is to be expected for an effect mediated by interfacial water molecules (Fig. 3b).

Calculations were also performed for positively charged particles assuming basic ionisable surface groups of p$K = 10$, giving $p = \text{p}H - \text{p}K = -4.5$. We find that the interaction remains monotonically repulsive, suggesting no cluster formation, which is consistent with the experimental observations (curves v-viii, Fig. 5c).

In conclusion, our findings show there may indeed be a plausible mechanism for the observed attraction of like-charged objects in aqueous solution. Rather than pointing to a failure of mean field theory, the experimental observations indicate the need for additional molecular level information absent in continuum theories: more specifically, the orientation of interfacial solvent molecules as shown in this study. While the framework of classical electromagnetics, and possible corrections from e.g., fluctuation forces[20] or charge inversion[48], would not support a symmetry broken response to complete inversion of the sign of charge in the system, the proposed interfacial mechanism unambiguously does. Although our present study focuses on explaining experimental observations in low ionic strength solution, the scaling of the screened electrostatic interaction implies that the same considerations could hold at much higher ionic strengths and at correspondingly closer distances of approach between the interacting objects. In particular the proposed mechanism may be capable of explaining a p*H*-tunable affinity between negatively charged macromolecules that is repulsive at higher p*H* and turns attractive under more acidic conditions even though the molecules carry substantial net negative charge over the entire p*H* range of interest[49]. This behaviour is distinct from the fluctuation induced attraction anticipated for molecules close to their isoelectric points (or point of zero charge)[50,51] and may be relevant in a broad range of phenomena, such as biological phase segregation[49], crystallisation[52], histone-bound packaging of DNA in the nucleus, formation and dissolution of polyphosphate stress granules[53], phosphorylation-based modulation of molecular interactions, or indeed in any system involving interactions between entities of low net negative charge density. It is worth noting that the p*H* range of interest in this study, 5.5-7, is in fact the range utilised in biological organisms to effect the formation and dissolution of biomolecular condensates and intracellular phase separation[49,54] where the relevant charged groups have p*K*s in the range 3-4 as for the colloidal particles in this study. Although the results presented at this stage for comparatively macroscopic objects would not warrant quantitative predictions on interactions at the molecular scale, the generality of the mechanism raises the distinct possibility of relevance in this context. Furthermore, chemical details of the surface may play an important role especially for complex materials such as zwitterionic lipids and oxides like silica composed of different species of ionisable groups with widely different p*K*s, and hydrogen bonding capability. Importantly, however, our findings based on MD simulations involving non-polar and non-hydrogen-bonding walls suggest that the chemical nature of the surface is likely play a role subordinate to its electrical charge. Although our minimal model of surface-water interactions should not be expected to provide a quantitative description of interactions in all systems, it is remarkable that this simplified picture is capable of explaining hitherto unexplained observations in remarkable detail. Future experiments will further rigourously test the predictions of this model and refinements thereof. Our findings could point to a new fundamental understanding of the contribution of molecular water in interparticle and intermolecular interactions in solution.